\documentstyle[12pt,psfig,epsf]{article}
\begin{document}
\newcommand{\be}{\begin{equation}}
\newcommand{\ee}{\end{equation}}
\newcommand{\ba}{\begin{eqnarray}}
\newcommand{\ea}{\end{eqnarray}}
\newcommand{\bi}{\bibitem}
\begin{titlepage}
\hbox to \hsize{\large \hfil  IHEP 97-90}
\hbox to \hsize{\large \hfil  December 1997}
\hbox to \hsize{\hfil }
\hbox to \hsize{\hfil }
\hbox to \hsize{\hfil }
%\vfill
\large \bf
\begin{center}
QCD RUNNING COUPLING: FREEZING VERSUS ENHANCEMENT IN THE
INFRARED REGION\footnotemark 
\footnotetext{Talk presented at the XIIth International Workshop 
on High Energy Physics
and Quantum Field Theory, 
September 4 -- 10, 1997, Samara, Russia}
\end{center}
\vskip 1cm
\normalsize
\begin{center}
{\bf Aleksey I. Alekseev
}\\
{\small {\it Institute for High Energy Physics,
  Protvino, Moscow Region, 142284 Russia}\\
E-mail: alekseev@mx.ihep.su}
\end{center}
\vskip 1.5cm
\begin{abstract}
We discuss whether or not "freezing" of the QCD running coupling 
constant in the infrared region is consistent with the Schwinger -- 
Dyson (SD) equations. Since the consistency of the "freezing" was not
found, the conclusion is made that the "analytization" method does 
not catch an essential part of nonperturbative contributions.
Proceeding from the results on consistency of the infrared enhanced 
behaviour of the gluon propagator with SD equations,  the running 
coupling constant is modified taking into account the minimality 
principle for the nonperturbative contributions in the ultraviolet
region and convergence condition for the gluon condensate.
It is shown that the requirements of asymptotic freedom, analyticity,
confinement and the value of the gluon condensate are compatible in 
the framework of our approach. Possibilities to find an agreement of 
the enhanced behaviour of the running coupling constant with integral 
estimations in the infrared region are also discussed.
\end{abstract}
\vfill
\end{titlepage}

The phenomenon of asymptotic freedom~\cite{Gros} called forth 
an impressive 
success of perturbative QCD in the description of experimental data
plethora. However, there is  wide scope of phenomena which 
is intractable in the framework of 
perturbation theory. Nonperturbative effects 
modify the infrared behaviour of the quark and gluon Green`s
functions. 
With $q^2$ decreas the 
renormalization group improved one-loop
running coupling constant
\begin{equation}
\bar\alpha_s (q^2)=\frac{4\pi}{b_0\ln (q^2/\Lambda^2)}.\label{a2}
\end{equation}
($b_0=11 C_2/3-2N_f/3$) increases,
which may indicate a tendency of unlimited growth of the interaction 
at large distances, leading to a confinement of coloured objects.
However, at $q^2=\Lambda^2$ in~(\ref{a2}) the pole is 
present, 
which is nonphysical, at least, due to the fail of the 
perturbation 
theory,
and the account 
of nonperturbative effects becomes obligatory. 

In recent papers~\cite{Shir} the solution of the problem of a ghost 
pole
was proposed with the  condition of analyticity in $q^2$ being 
imposed.
The idea of "forced analyticity" goes back to~\cite{Red,Bog}
of the late 50s. They were dedicated to the problem of
Landau-Pomeranchuk pole~\cite{Land} in QED. Using for 
$\bar\alpha_s(q^2)$
a spectral representation without subtractions, the following 
expression
for the running coupling constant was obtained in~\cite{Shir}
\be
\bar\alpha_s^{(1)}(q^2)=\frac{4\pi}{b_0}\left [
\frac{1}{\ln(q^2/\Lambda^2)} +\frac{\Lambda^2}{\Lambda^2-q^2}\right ].
\label{a3}
\ee
This expression 
is analytic in the infrared region  due to nonperturbative
contributions
and it has a finite limit at zero (although the derivative is infinite).

Nowadays a possibility of "freezing" the coupling constant
at low energies is under discussion~\cite{Mat} in the framework of some
scheme of approximate calculations ($(16\frac{1}{2}- N_f)$ - expansion).
In the approach~\cite{Sim} with confining background field,
"freezing" was also obtained,
\begin{equation}
\bar\alpha_s (q^2)=\frac{4\pi}{b_0\ln ((m^2_B+q^2)/\Lambda^2)}.\label{s1}
\end{equation}
Here $m_B$ is a process-dependent constant of the order of 1 GeV.

In the present paper 
we follow  the approach  of
%which is essentially based on 
Refs.~\cite{Arb,prep}.
We discuss the problem of consistency of the
constant behaviour of the running coupling constant 
in the infrared region 
with Schwinger-Dyson (SD) equation for a gluon 
propagator. Further we include into consideration 
additional nonperturbative
terms, in particular, the singular
in the infrared region term $\sim 1/q^2$, the necessity 
of renormalization invariance being taken into account. Then we 
consider
possibilities to fulfill the
demands of confinement, asymptotic 
freedom,
analyticity,  
correspondence with estimates of the gluon condensate 
and integral estimates for $\bar\alpha_s(q^2)$
in the infrared region.

Let us consider the integral 
SD  equation for the gluon propagator in ghost-free 
axial gauge~\cite{Kum}
$A^a_\mu\eta_\mu=0,\;\; \eta_\mu$ --- gauge vector, $\eta^2\not=0$.
In this gauge the running coupling constant is directly connected with the
gluon propagator and Slavnov-Taylor identities \cite{Slav} have
the  simplest form. 
An important preference of the axial gauge consists in a
possibility to exclude the term from the SD equation, which
contains a full four-gluon vertex, by means of contraction
of the equation with tensor $\eta_\mu\eta_\nu/\eta^2$.
We  work in the Euclidean momentum space,
where smallness of the momentum squared 
means
smallness of all its components. 
For the gluon propagator
$D_{\mu\nu}(p)$, 
we suppose the approximation $D_{\mu\nu}(p) = Z(p^2) D_{(0)\mu\nu}
(p)$ to be appropriate for studying the infrared region. 
$D_{(0)\mu\nu}(p)$ is a free gluon propagator.
In the first paper of Ref.~\cite{prep} 
the possibility of the infrared behaviour
for renormalized function 
$Z_R(p^2)=Z(p^2)/Z(\mu^2)$ being of the form
\be
Z_R(x)=Z_R(0)+o(1),\;\; x\to +0,\label{a14}
\ee
has been studied ($Z_R(0)$ is nonzero constant)
in the framework of nonperturbative approach of Baker - Ball - Zachariasen
(BBZ)~\cite{Baker}.
The approximations of the BBZ approach as well as 
the condition $y=0$ imposed on gauge parameter $y=(p\eta)^2/p^2\eta^2$
seems to be adequate to studying the possibility of the infrared behaviour
which is not too singular.
With the assumption~(\ref{a14})  the equation for function
$Z_R(p^2)$ takes the form
\be
Z^{-1}_R(p^2)=1+Z_R(0)\frac{g^2C_2}{16\pi^2}\frac{11}{3}\ln p^2+o(\ln p^2),
p^2\to 0. \label{a33}
\ee
We see, that the behaviour $Z_R(p^2)\simeq Z(0)\not=0$ for $p^2\to 0$
does not agree with the SD equation. 

\begin{figure}[thb]
\centerline{\psfig{figure=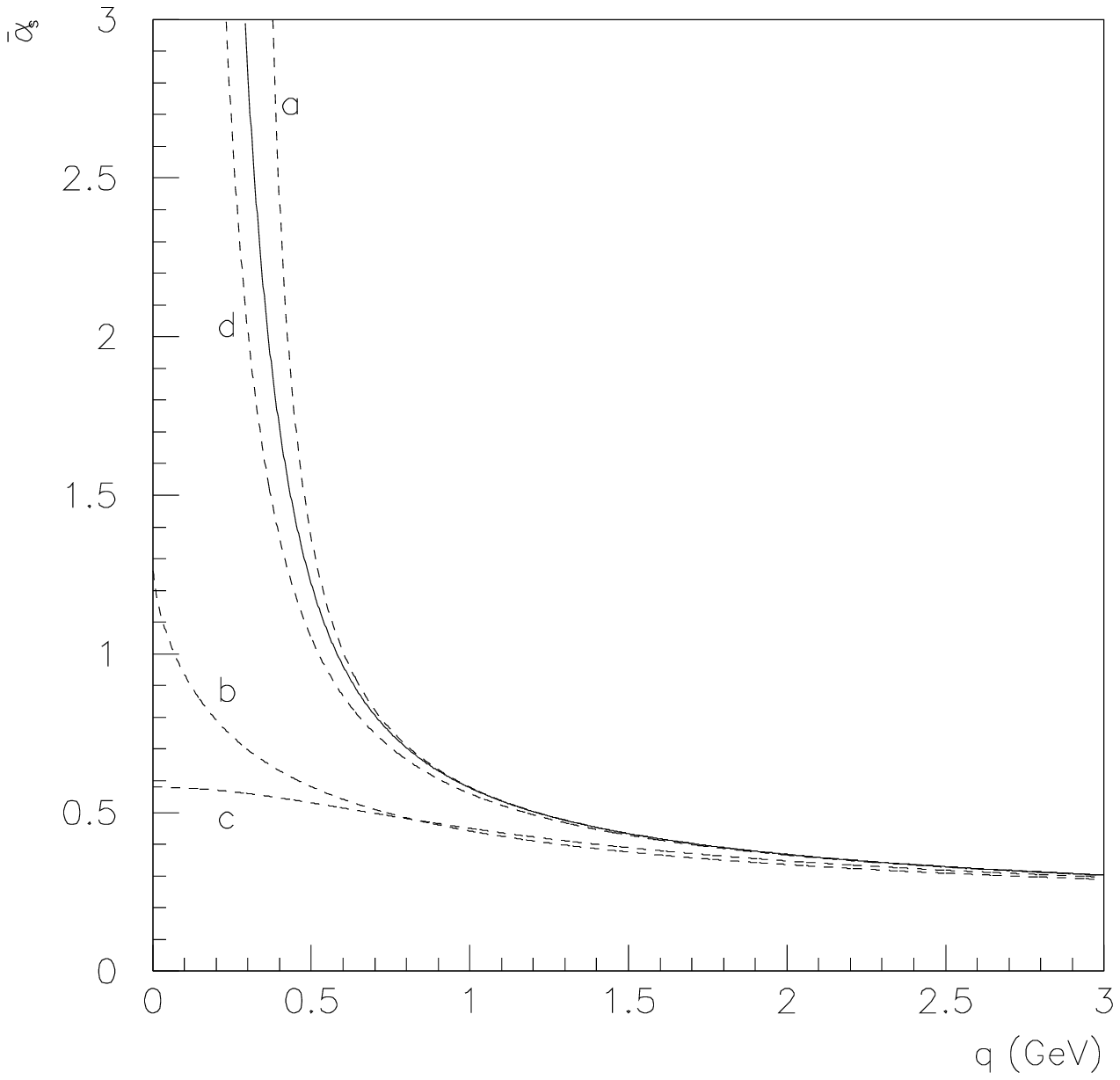,height=12cm,width=12cm}}
\caption{\label{samfig1} 
Momentum behaviour of the running coupling constant. Continuous curve  
corresponds to our Eq.~(7); the curve (a) is one-loop Eq.~(1);
the curve (b) shows "analytized" Eq.~(2); the curve (c) is "freezed"
behaviour Eq.~(3) where $m$ is taken $1$ GeV;
the curve (d) corresponds to Richardson parameterization 
for the running coupling constant (Eq.~(3)
with $m_B =\Lambda$). Here $\Lambda =300$ MeV, $b_0=9$.}
\end{figure}

This conclusion encourages  us to look for 
the possibilities different from the assumption  on the finiteness 
of the coupling constant at zero. Recently a possibility of the 
soft singular power infrared behaviour of the gluon propagator has
been discussed~\cite{Cud}, 
$D(q) \sim (q^2)^{-\beta}$, $q^2 \rightarrow 0$, 
where $\beta$ is a small positive non-integer number. In 
Ref.~\cite{AlekPL} the consistency of such behaviour
with the same equation was studied. A characteristic equation for the 
exponent $\beta$ was obtained and this equation was shown  to 
have no solutions in the region $0 < \beta < 1$. The authors of 
Ref.~\cite{Butner} also came to the conclusion on the inconsistency 
of the soft singular infrared behaviour of
the gluon propagator. The case of possible interference of power 
terms was studied in Ref.~\cite{AkekTMF96} and it was shown that in 
a rather wide interval $-1 < \beta < 3$ of the non-integer values 
of the exponent the characteristic equation had no solutions
\footnote{Note that this interval contains values $-1 < \beta < 0$
for which the propagator vanishes at zero.}. 
At present
a more singular, in comparison with free case, infrared behaviour 
of the form $D(q) \simeq M^2/(q^2)^{2}$, $q^2 \to 0$ seems to be 
most justified~\cite{Pagels,Baker,Alek1}. 
The models based on this assumption 
are widespread and rather attractive. In this way
it is possible to describe the dynamical chiral symmetry
breaking, to solve the $U(1)$ problem, to evaluate the
topological susceptibility~\cite{1,2}, to calculate
the condensates of gluon and quark fields~\cite{3}, etc.
The physical consequences of such enhancement of zero modes are 
discussed in  reviews~\cite{ArbPartNucl,RobWil}. 
Bearing in mind the remarks stated above, let us 
turn to the problem of nonperturbative contributions.
The "analytized" expression~(\ref{a3}) without nonphysical
singularity will be a starting point of our further consideration.
We see that this expression has a nonperturbative tail with the
behaviour $1/q^2$ at $q^2 \to 0$. To answer the question if
this behaviour is admissible,  let us consider
the important physical quantity, namely, the gluon condensate
$K\,=\,{<vac\mid \alpha_s/\pi:G^a_{\mu \nu}\,G^a_{\mu \nu}:\mid 
vac>}\,$. 
According to the definition (see e.g.,~\cite{ArbPartNucl}) 
up to the quadratic approximation in the gluon fields, one has
after the Wick rotation
\begin{eqnarray}
K\,=\,-\,\frac{\delta^{aa}}{\pi}\int\frac{d^4k}{(2\pi)^4}
(\delta_{\mu \nu}k^2 - k_\mu k_\nu)D^{(0)}_{\mu \nu}(k)\frac{g^2}
{2\pi}(Z(k) - Z^{pert}(k)) =  \nonumber \\
=\,\frac{48}{\pi}\,\int \frac{d^4k}{(2 \pi)^4}\,
\left(\bar\alpha_s(k^2)\, - \,\bar\alpha_s^{pert}(k^2)\right)\,=\,
\frac{3}{ \pi^3} \,\int_0^{\infty}\,\bar\alpha_s^{np}(y)\,y dy\, ,
\label{a44}
\end{eqnarray}
where $\bar\alpha_s^{np}$ is a nonperturbative part of the 
running coupling constant. 
The one-loop "analytized" behaviour of Eq.~(\ref{a3}) leads to the
quadratic divergence in Eq.~(\ref{a44}) 
at infinity
and this is true for
two- and three-loop expressions~\cite{Shir} of the analytization 
approach. Note that at large $q^2$, "freezed" behaviour~(\ref{s1})
and known enhanced in the infrared region Richardson's
running coupling constant~\cite{Rich}
(coinciding formally with~(\ref{s1}) at $m_B=\Lambda$)
do not ensure the convergence of the integral~(\ref{a44})
at infinity.
According to the results of Refs.~\cite{Pagels,Baker,Alek1}
let us add in Eq.~(\ref{a3}) the isolated infrared singular term
of the form $1/q^2$. This term is harmless at zero and it can improve
the behaviour of the integrand at infinity and make the integral
logarithmically divergent. To make integral~(\ref{a44}) convergent
at infinity, it is sufficient to add one more isolated singular term
of a pole type with parameters chosen appropriately.
In this sense the model we come to is minimal.
The expression, we obtain for the running coupling constant, is the
following:
\be      
\bar\alpha_s(q^2)\,=\,\frac{4 \pi}{b_0}\Biggl(\frac{1}{
ln(q^2/\Lambda^2)}\,
+\,\frac{\Lambda^2}{\Lambda^2 - q^2}\,+\,\frac{c \Lambda^2}{q^2}\,+
\,\frac{(1-c) \Lambda^2}{q^2 + m_g^2}\Biggr)\,,\label{a45}
\ee
with  mass parameter $m_g$,
\be
m_g^2\,=\,\Lambda^2/(c-1).\label{a46}
\ee
It is worth  noting that an account of nonperturbative contributions
in Eq.~(\ref{a45}) preserves a perturbative time-like cut of 
Eq.~(\ref{a3}).
With the given value of the QCD scale 
parameter $\Lambda$, the parameter
$c$ can be fixed by the string tension $\kappa$ or the Regge slope 
$\alpha'=1/(2\pi \kappa)$ assuming the linear confinement
$V(r)\simeq \kappa r=a^2r$ at $r\to \infty$. 
We define the potential $V(r)$ of static $q\bar q$ interaction
~\cite{BoShir,Buch} by means of three-dimensional Fourier 
transform of $\bar \alpha_s(\vec q^2)/\vec q^2$ with 
the contributions of only one dressed gluon exchange taken 
into account. This gives the following relation
\be
c\Lambda^2=(3b_0/8\pi)a^2=(b_0/16\pi^2)g^2M^2. \label{aa36}
\ee
Taking  $a\simeq 0.42\,GeV$, one obtains $c=\Lambda^2_1/\Lambda^2$
where $\Lambda^2_1= 3b_0 \kappa / 8\pi\, \simeq 0.434\, GeV$ ($b_0 = 9$ in
the case of 3 light flavours). 
From Eq.~(\ref{a46})  one obtains 
\be
m_g\,=\,\frac{\Lambda^2}{\sqrt{\frac{27}{8\pi}a^2 -\Lambda^2}},\label{s46}
\ee
and the tachion absence condition limits the parameter $\Lambda$,
$\Lambda < 434 MeV$.
With Eq.~(\ref{aa36}) taken into account, the  parameters
in Eq.~(\ref{a45}) are only $\Lambda$ and $b_0$.
For $SU_c(3)$ colour group $(C_2=3)$, $N_f=3$ and $\Lambda=300$ MeV,
the behaviour of running coupling constant is represented in
Fig.~\ref{samfig1} for the cases under discussion.
%
%                 Fig.1
It should be noted that at sufficiently large $q$ the curves
in Fig.~\ref{samfig1} seem to be indistinguishable, but
the presence of amplifying multipler $y$ in Eq.~(\ref{a44})
allows one to distinguish the ultraviolet behaviour of
the nonperturbative contributions.

Let us represent expression~(\ref{a45}) in explicitly 
renormalization 
invariant form. It can be done without solving the differential 
renormalization group equations. In this order we write
$\bar\alpha_s(q^2)=\bar g^2(q^2/\mu^2,\; g^2)/4\pi$ and use the 
normalization condition $\bar g^2(1,g^2)=g^2$. Then we obtain
the equation for the wanted dependence of the parameter $\Lambda^2$ on
$g^2$ and $\mu^2$:
$$
g^2/4\pi=\frac{4\pi}{b_0}
\left [
\frac{1}{\ln (\mu^2/\Lambda^2)}+\frac{\Lambda^2}{\Lambda^2-\mu^2}+c
\frac{\Lambda^2}{\mu^2}
+\frac{(1-c)\Lambda^2}{\mu^2+m^2_g}
\right ].
$$
From dimensional reasons
$
\Lambda^2=\mu^2 exp\{-\varphi (x)\}$,
where $x=b_0g^2/16\pi^2=b_0\alpha_s/4\pi$, and for function
$\varphi (x)$ we obtain the equation:
$$
x=\frac{1}{\varphi (x)}+\frac{1}{1-e^{\varphi (x)}}+ce^{-\varphi 
(x)} - \frac{(c-1)^2}{(c-1)e^{\varphi (x)}+1}.
$$
The solution of this equation at $c>1$  is  
function $\varphi (x)$, which has the behaviour $\varphi (x)\simeq 
1/x$ at $x\to 0$ and  $\varphi (x)\simeq -\ln(x/c)$ at
$x\to +\infty$. 
The relation obtained ensures the renormalization invariance of
$\bar\alpha_s(q^2)$. At low  $g^2$, we obtain
$
\Lambda^2=\mu^2\exp\{-4\pi/(b_0\alpha_s)\}$,
which indicates the essentially nonperturbative character of
three last terms in Eq.~(\ref{a45}) and these terms are absent
in the usual perturbation theory.

\begin{figure}[thb]
\centerline{\psfig{figure=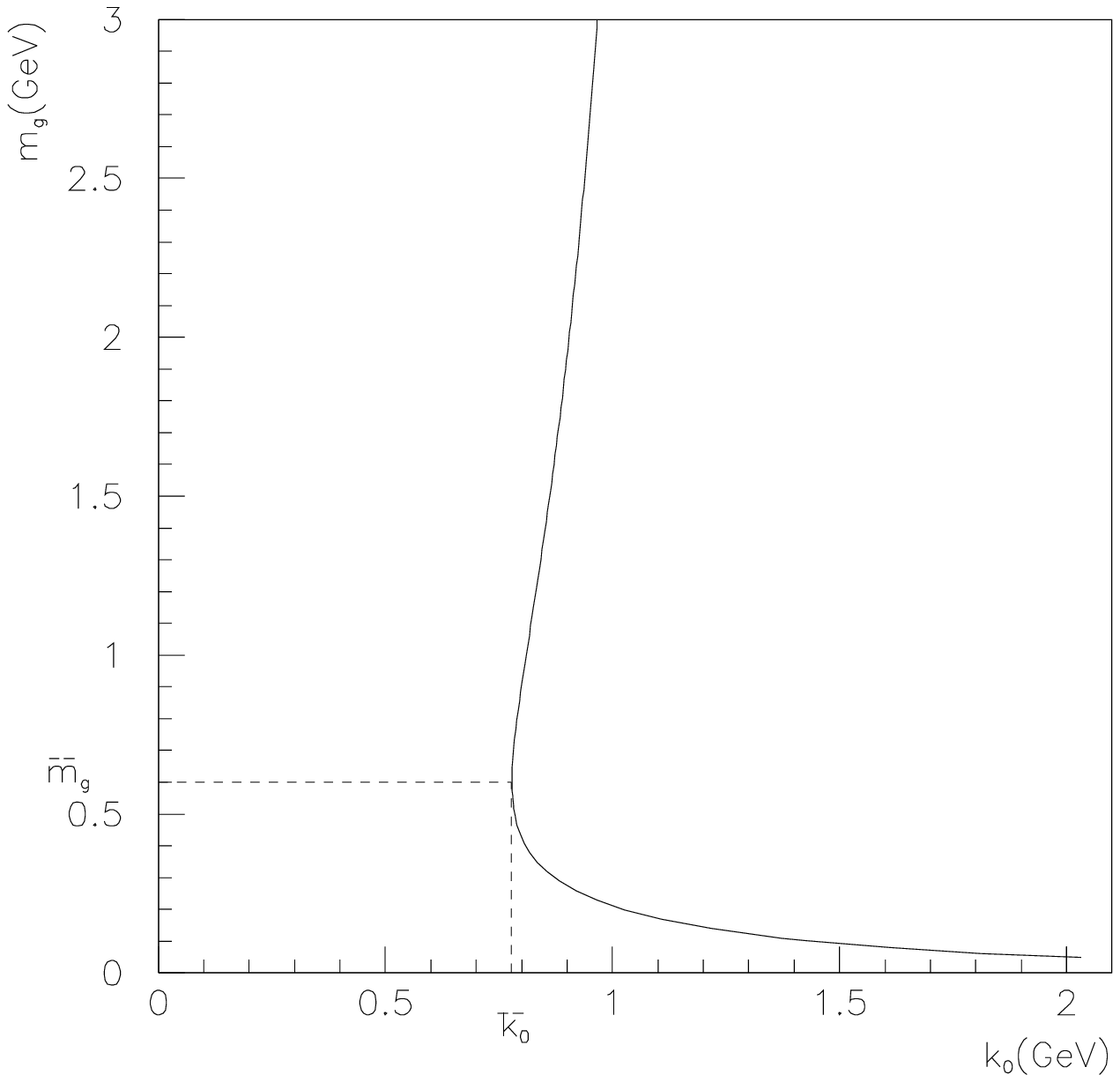,height=12cm,width=12cm}}
\caption{\label{samfig2} 
The dynamical gluon mass $m_g$ corresponding to the nonperturbative 
scale  
$k_0$ for fixed value of the gluon condensate
$K^{1/4}=0.33 $ GeV. }
\end{figure}

The acceptance of the cancellation mechanism for the nonphysical
perturbation theory singularity in Eq.~(\ref{a2}) by the nonperturbative 
contributions leads to the necessity of supplementary definition of
integral~(\ref{a44}) near point  $k^2 = \Lambda^2$.
This problem can be reformulated as a problem of dividing
perturbative and nonperturbative contributions in $\bar\alpha_s$
resulting in the introduction of some parameter $k_0 \approx 1$ GeV.
This provides the absence of the pole at $k^2=\Lambda^2$ in 
both perturbative and nonperturbative parts
\footnote{See also Ref.~\cite{Grun} where the problem of perturbative 
and nonperturbative contributions to $\bar\alpha_s$ is discussed
and the definition of infrared finite regularized perturbative
part of $\bar\alpha_s$ is suggested.}.
Nonperturbative contributions in Eq.~(\ref{a45}) decrease at 
infinity as $1/q^6$, the integral in Eq.~(\ref{a44}) converges and 
we can obtain
\be
K(\Lambda,k_0)\,=\,\frac{4}{3 \pi^2}\,\left\{\Lambda^4\,ln\left[
\left(\frac{\Lambda^2_1}{\Lambda^2}-1\right)\left(\frac{k^2_0}{\Lambda^2}
-1\right)\right] +k^2_0\Lambda^2\right\}
\,.\label{a47}
\ee
Phenomenology gives the positive value of the gluon condensate $K$ 
in the interval  $(0.32\,GeV)^4$ --- 
$(0.38\,GeV)^4$~\cite{Vain,Grein}. 
If one takes the value of gluon condensate $K^{1/4} = 0.33 GeV$
and consider formula~(\ref{a47}) 
as equation for $\Lambda $
at different values of $k_0$, the picture will be the following.
At $k_0 < \bar k_0 =0.777 \,GeV$ there are no solutions in the
interval
$\Lambda = 0.1\, GeV - 0.434\, GeV$.
At $k_0 = \bar k_0$ there is a single solution $\Lambda = 375 MeV$
corresponding to $m_g =\bar m_g \simeq 0.6\, GeV$. 
At $k_0 > \bar k_0$ for $\Lambda$  two solutions appear
to whom two values 
$m_g$ correspond, one of them increases with the increase of $k_0$ 
and other decreases with increase of $k_0$.
This situation is illustrated in Fig.~\ref{samfig2} where we
used Eq.~(\ref{s46}) to connect the parameters $\Lambda$
and $m_g$.
%
%                 Fig.2
%

It is seen from Eq.~(\ref{a45}) that the pole singularities are
situated at two points $q^2 = 0$ and $q^2 = - m_g^2$.
It corresponds to the two effective gluon masses, $0$ and $\,m_g$. 
Therefore, the physical meaning of the parameter $m_g$ is not
the constituent gluon mass, but rather the mass of the exited state 
of the gluon. 

In a number of cases of the QCD calculations it is necessary to
estimate integrals of the form
\be
F(q^2)=\int\limits^{q^2}_{0}\bar\alpha_s(k^2)f(k^2)dk,\label{a37}
\ee
where $f(k^2)$ is some smooth function and integration includes 
the infrared region where the usual perturbation theory is 
inapplicable.
In this cases the infrared matching scale  $\mu_I$ 
can be introduced $ (\Lambda <<\mu_I<<q)$
and the contribution to integral~(\ref{a37})
from the region $k>\mu_I$ can be calculated perturbatively.
To estimate the contribution of the region  $k<\mu_I$ 
in Ref.~\cite{Dok} the running coupling constant is assumed to be
regular at zero and power terms are extracted using dimensional 
arguments. In this connection the nonperturbative parameters $\alpha_p$
are introduced,
\be
\alpha_p(\mu_I)=\frac{p+1}{\mu_I^{p+1}}\int\limits^{\mu_I}_{0}
dk\bar\alpha_s(k)k^p,\label{a38}
\ee
which should be found from experiment.
In the approach connected with infrared renormalons~\cite{Zakhar},
the analogous ambiguity arises when evaluating 
integral~(\ref{a37}). 
This ambiguity can be understood in the following way.
Expanding $\bar\alpha_s(k^2)$ in the series of powers
of $\bar\alpha_s(q^2)$ and integrating in $k^2$ one obtains
an asymptotic series with terms growing as factorial 
which can be estimated by the finite number of terms up to
power corrections.
The characteristic integral  $\bar A(\mu) \equiv \alpha_0(\mu)/\pi$
(zero moment) has been estimated~\cite{Dok,Dok1} 
with the result
$\alpha_0(2GeV)/\pi = 0.18 \pm 0.02$.
This integral turned out to be not only stable with respect
to the choice of the infrared regularization (fit-invariance)
but also relative to different approximations of the high energy
tail of the running coupling (number of active quarks, one- or
two-loop approximation). It is shown in Refs.~\cite{Shir} 
that the behaviour~(\ref{a3}) complies with integral estimations 
of Refs.~\cite{Dok,Dok1}.
Let us see whether the behaviour~(\ref{a45})
is compatible with these estimations.
In this case we also encounter the problem of definition in expressions
of form~(\ref{a37}) and the necessity to fix the corresponding 
ambiguity. The following method seems to be highly convenient
and universal,
\be
\frac{1}{(k^2)^m} \Longrightarrow \lim_{\epsilon'\to 0}\frac{d}{d\epsilon'}
\frac{\epsilon'}{(k^2)^m}\left(\frac{\mu^2}{k^2}\right)^{\epsilon'}.
\label{a39}
\ee
The finite arbitrariness described by the constant $\mu^2$
arises only if we integrate the logarithmic singularities
(which are usually connected with physics). In the case of
local integrability ($n > 2m$, here $n$ is the dimension of integration
space) the rule~(\ref{a39}) runs idle and at $n < 2m$ 
it corresponds~\cite{Gelfand} to analytic continuation in $m$. 
In accordance with the rule~(\ref{a39}),
the contribution of the singular term of Eq.~(\ref{a45}) to the 
moments~(\ref{a38}) is the following:
\be
\Delta^{sing}\alpha_p(\mu_I) = \frac{p+1}{p-1}\frac{3a^2}{2\mu_I^2},
p \ne 1,\; \Delta^{sing}\alpha_1(\mu_I) =-\frac{3a^2}{2\mu_I^2}
\ln\frac{\mu^2}{\mu_I^2}.\label{a40}
\ee 
Thus, at $p \le1$ the singular term of Eq.~(\ref{a45}) should be
considered as a distribution and fit-invariance of the lowest moment
$\bar\alpha_0(\mu_I)$
for different variants of infrared behaviour can probably indicate
the absence of "freezing"  the running coupling constant and
universality of ambiguity fixing. In favour of "freezing"  the 
coupling could indicate the equality of moments  
$\alpha_p(\mu_I)$ for different $p$. 
For contribution of the last two terms of Eq.~(\ref{a45})
to zero moment one obtains:
\be
\Delta\alpha_0(\mu_I)/\pi = -\frac{4}{9}\frac{c\Lambda^2}{\mu^2_I}
-\frac{4}{9}\frac{\Lambda}{\mu_I}(c-1)^{3/2} arctg(\frac{\mu_I}{\Lambda}
\sqrt{c-1}). \label{s2}
\ee
Taking $a \simeq 0.42 GeV$, $\mu_I = 2 GeV$  one finds rather small
contribution of the singular term, 
$\Delta^{sing}\alpha_0(\mu_I)/\pi $ $= -(3a^2)/(2\pi\mu_I^2)
\simeq- 0.021$.
At $\Lambda \rightarrow \Lambda_1$ the second term of Eq.~(\ref{s2}) 
vanishes as $-(16/9)(\Lambda_1 - \Lambda)^2/\Lambda^2_1$ and it can 
be made small by a corresponding choice of  $\Lambda$.
For example, at $\Lambda = 0.375 GeV$ (in this case $c= 1.3476$)
this term equals  $- 0.0216$. 
Having in mind the results of Refs.~\cite{Shir} 
where the contribution of the first two terms of Eq.~(\ref{a45})
has been evaluated, we conclude that the running coupling constant
Eq.~(\ref{a45}) can be consistent with integral estimations in the 
infrared region.
Note, that if we compare Eq.~(5) and Eq.~(12) of the second of 
Ref.~\cite{Shir},
we find that "analytization" procedure in two-loop case leads, at 
large momentum, to the half nonperturbative contribution of 
one-loop case. It can point to  the tendency of minimization
of nonperturbative contributions in the ultraviolet region.

I would like to thank B.A.~Arbuzov and V.E.~Rochev for interesting 
discussions. This work is supported by the Russian Foundation for
Basic Research
Grant 95-02-03704.
%
%            REFERENCES

\end{document}